\documentclass[aps,pre, twocolumn,superscriptaddress]{revtex4}
\usepackage{graphicx}
\usepackage{color}

\begin{document}


\title{XUV Opacity of Aluminum between the Cold-Solid to Warm-Plasma Transition}


\author{S.M.Vinko}
\affiliation{Department of Physics, Clarendon Laboratory, University of Oxford, Parks Road, Oxford, OX1 3PU, UK}

\author{G. Gregori}
\affiliation{Department of Physics, Clarendon Laboratory, University of Oxford, Parks Road, Oxford, OX1 3PU, UK}

\author{B. Nagler}
\affiliation{Department of Physics, Clarendon Laboratory, University of Oxford, Parks Road, Oxford, OX1 3PU, UK}

\author{T.J. Whitcher}
\affiliation{Department of Physics, Clarendon Laboratory, University of Oxford, Parks Road, Oxford, OX1 3PU, UK}

\author{M.P. Desjarlais}
\affiliation{Pulsed Power Sciences Center, Sandia National Laboratories, Albuquerque, New Mexico 87185, USA}

\author{R.W. Lee}
\affiliation{Lawrence Livermore National Laboratory, University of California, P.O. Box 808, CA 94551, USA}

\author{P. Audebert}
\affiliation{LULI, \'Ecole Polytechnique, CNRS, CEA, UPMC, Route de Saclay, 91128 Palaiseau, France}

\author{J.S. Wark}
\affiliation{Department of Physics, Clarendon Laboratory, University of Oxford, Parks Road, Oxford, OX1 3PU, UK}


\date{\today}

\begin{abstract}
We present calculations of the free-free XUV opacity of warm, solid-density aluminum at photon energies between the plasma frequency at 15 eV and the L-edge at 73 eV, using both density functional theory combined with molecular dynamics and a semi-analytical model in the RPA framework with the inclusion of local field corrections. As the temperature is increased from room temperature to 10 eV, with the ion and electron temperatures equal, we calculate an increase in the opacity in the range over which the degree of ionization is constant. The effect is less pronounced if only the electron temperature is allowed to increase. The physical significance of these increases is discussed in terms of intense XUV-laser matter interactions on both femtosecond and picosecond time-scales.
\end{abstract}

\maketitle

The change in the opacity of material in the XUV regime as it makes the transition from a cold solid to a warm plasma is a problem that both tests our fundamental understanding of how light interacts with matter and is one of increasing experimental interest due to the development of ultra-intense XUV free-electron-lasers (FELs). For example, the FLASH XUV-FEL based in Hamburg now produces 15-fs pulses of XUV radiation tunable between 30 and 6 nm, with photon numbers $\ge 10^{12}$ per bunch, i.e., peak powers in the GW range, at 10 Hz. When focussed with multi-layer-coated off-axis parabolic mirrors to spot diameters of order a few microns, irradiances in excess of 10$^{16}$ Wcm$^{-2}$ can be produced - an irradiance regime that has until  recently remained the province of optical lasers. A thin foil placed in the focus of such a beam would be rapidly heated to several tens of eV, and during this absorption process we expect significant changes in the opacity before the onset of any hydrodynamic motion and consequent density change. Therefore, to have an understanding of the femtosecond heating and provide a predictive capability for experimental design, a consistent absorption model is required. A complete model, which covers a wide range of densities, temperatures, and incident photon energies  would need to include sophisticated atomic physics that accounted for both bound-free and free-free absorption, and how these varied with charge state upon heating and ionization.  The ability to separate the contributions of these different mechanisms would provide the opportunity to test our ability to model the role that each has to play in the opacity of  dense plasmas.

It is in this context that the opacity of warm dense Al in the XUV regime is of particular interest.  
With a Fermi energy of order 11 eV,  plasmon energy just above 15 eV and an L-edge at 73 eV, there is a large extent of energy space where free-electron theory might be thought to be applicable. Indeed, for solid Al, band calculations of the density of states (DOS) are in excellent agreement with free-electron theory up to the pseudogap between 3d and 4f bands, at around 40 eV~\cite{Takada:2002p216402}.
Importantly, calculations of the equation of state indicate that for solid density Al in thermal equilibrium, ionization of the L-shell does not start to occur until electron temperatures close to the Fermi temperature~\cite{Kim:2003p56410}.  Thus, with a known free-electron density and ionic state over a wide temperature range, it is an ideal material to test our understanding of free-free absorption in warm dense matter (WDM). 

In this rapid communication we present calculations for solid density Al at temperatures of up to 10 eV using a semi-analytical model for the dielectric response within the random phase approximation (RPA) framework and with inclusion of particle-hole interactions in terms of a local field correction (LFC). We test the validity of the model by conducting {\it ab initio} molecular dynamics (DFT-MD) calculations using the VASP code~\cite{Kresse:group} in conjunction with a Kubo-Greenwood (KG) calculation of the optical properties~\cite{Desjarlais:2002p25401,Mazevet:2005p16409}. A steady rise in opacity as a function of temperature for solid density Al for photon energies between the plasma frequency and about 40 eV is calculated, which is sufficiently large to be experimentally verifiable. We conclude by commenting on the remaining discrepancies between calculated and observed XUV opacities for solid-state, room-temperature Al.

The absorption of light of angular frequency $\omega$ by a material having a macroscopic dielectric function $\epsilon_M$ is given in terms of the absorption coefficient as
$
\kappa = 2 \omega c^{-1} \mathrm{Im} \sqrt{ \epsilon_M (\omega)}.
$
For a quasi-free electron metal the macroscopic dielectric function can be expanded in a perturbation series of the potential between electrons and ions under the assumption that this is small compared to other energies of the system. Up to second order, this expansion has the form (see, {\em e.g.}, Ref.~\cite{Sturm:1982p1})
$
\epsilon_M (\omega) = 1 - \omega_p^2/\omega^2 + \epsilon^{(2)}(\omega).
$
Here $\omega_p$ denotes the electron plasma frequency $\omega_p^2 = n_e e^2 /( \epsilon_0 m_e)$, where $n_e$ and $m_e$ are the electron density and mass.
A systematic approach to the calculation of $\epsilon^{(2)}(\omega)$, the first term describing the modulation of the free-electron gas due to the presence of the ions, was first conducted by Ron and Tzoar~\cite{Ron:1963p12}, who's result can be written in the form:
\begin{eqnarray} \label{Eq:ron-tzoar1}
\epsilon^{(2)}(\omega) = \frac{n_i}{6 \pi^2 m^2_e \omega^4}\int dq q^6 \frac{V_q^2}{|\varepsilon(q, \omega)|^2} S(q)
\left [ \varepsilon(q, \omega) - \varepsilon(q, 0) \right ].
\end{eqnarray}
Here $V_q$ indicates the Fourier transform of the electron-ion interaction potential, $S(q)$ the ion-ion structure factor and $\varepsilon(q, \omega)$ the electronic microscopic dielectric function. This expression has been shown to be exact to second order in the potential~\cite{Hopfield:1965p1401}. We note that Eq.~(\ref{Eq:ron-tzoar1}) corresponds to the Balescu-Lenard approximation in the kinetic theory approach~\cite{Gould:1967p68}. Further, in the context of the requirement for a theory that is applicable to both the cold-solid and hot low-density plasma extremes, we note that it is a known result that Eq.~(\ref{Eq:ron-tzoar1})  is well suited to reproduce the classical plasma limit, {\it i.e.} inverse bremsstrahlung, when the Coulomb potential and classical forms of structure factors are considered~\cite{Dawson:1962p517}. The applicability of Eq.~(\ref{Eq:ron-tzoar1}) is based on the assumption of quasi free-electron structure and can be as such easily extended to a range of simple metals.

Within the RPA, the electron dielectric function can be evaluated as a function of temperature throughout the degenerate to non-degenerate regime~\cite{Arista:1984p1471}. The RPA is exact in the high density limit, where long range correlations dominate. At lower densities, such as those in real metals, intermediate and short range correlations become increasingly important and are described by vertex and self-energy corrections. A case of particular interest are electron-hole ladder interactions describing excitons, which have been shown to give large contributions to the absorption strength~\cite{Takada:2002p216402}. We add such contributions to the RPA dielectric function in terms of a LFC following the work of Higuchi and Yasuhara~\cite{Higuchi:2000p2099}, including full temperature dependence to the RPA polarisation functions and calculating the averaged electron screening over a finite-temperature Fermi sphere. While this approximation might not be sufficient for an in detail study of many-body contributions, it does capture the main effect in Al which is an increase in the free-free absorption of about a factor two in the range of interest.

In evaluating the ion-ion structure factor in Eq.~(\ref{Eq:ron-tzoar1}), there are different approaches that we can use for WDM. Numerically, molecular dynamics (MD)~\cite{Hansen:2006} techniques as well as direct solution of the hyper-netted chain (HNC) equations~\cite{Ichimaru:2004,Hansen:2006} can be implemented. While accurate for ideal classical plasmas, both MD and HNC become computationally expensive if the full dynamic response of the quantal electron background has to be accounted for. In order to simplify the analysis,  we use the approach initially suggested by Singh and Holz~\cite{Singh:1983p1108} for a liquid metal and applied to the ion-ion structure factor of WDM by Gregori {\it et al.}~\cite{Gregori:2007p99}. In this approach, the bare ionic response is constructed from the analytical solution of the mean spherical equations for a system of charged hard spheres embedded in a neutralizing background~\cite{Palmer:2003p4171}. Such an approximation is the equivalent of the Percus-Yevick solution for a neutral gas of hard spheres~\cite{Waisman:1972p3086,Hansen:2006}, and has been shown to correctly reproduce MD and HNC results at various degrees of inter-ion coupling~\cite{Hansen:2006}. The idealized  one-component plasma (OCP) structure factor $S_{\mathrm{OCP}}(q)$ is thus obtained in a fully analytical form. The dynamical screening of the electron background on the ions is then calculated within the linear response framework~\cite{Galam:1976p816,Gregori:2007p99}:
$
S_{\mathrm{WDM}}(q) =  S_{\mathrm{OCP}}(q)/ [1 + f_v(q) S_{\mathrm{OCP}}(q)],
$
where $f_v(q)$ is the attractive screening correction to the bare ion-ion interaction~\cite{Singh:1983p1108,Gregori:2007p99}. The agreement of this model with the structure factor computed by DFT-MD is seen to be very good.

\begin{figure}
\includegraphics[width=\linewidth]{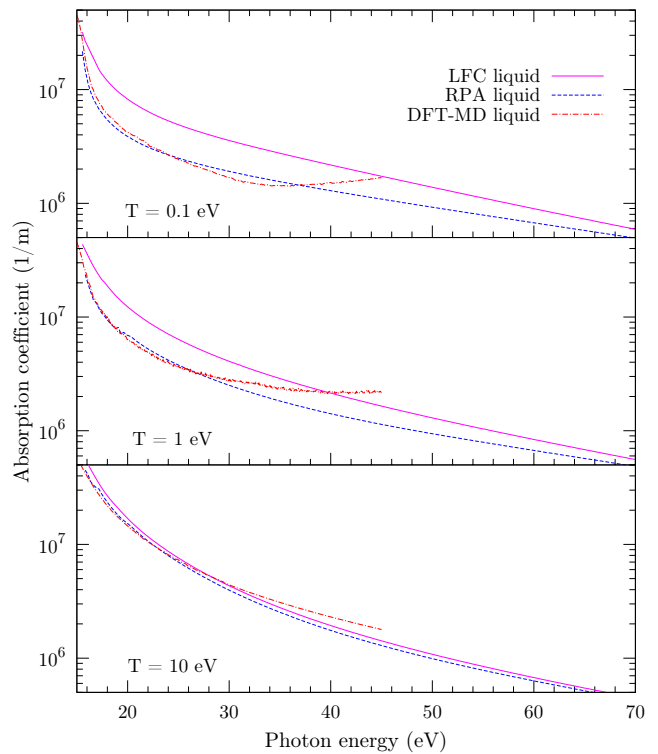}
\caption{\label{Fig:frequency} (color online) Absorption coefficient versus photon energy for solid density Al, Z=3. Comparison between results from Eq.~(\ref{Eq:ron-tzoar1}) in the RPA (blue) and with electron-hole LFC (purple) and DFT-MD simulations (red).}
\end{figure}

The Coulomb interaction potential is not adequate to describe the degenerate regime, where the ionic core potential is both screened by the electrons and modified by the interaction with nearby ions. To take these effects into account we use the empty-core pseudopotential model, where the potential is zero inside a certain distance from the ion core, $r_c$, and  $-Ze^2 / (4 \pi \epsilon_0 r)$ outside. Its Fourier transform is given by
$
V_q = -Z e^2\cos(qr_c)/ (\epsilon_0 q^2).
$
Here we use $r_c = 0.60 \mathrm{\AA}$, which minimizes the difference between the pseudopotential values and experimental data obtained through measurements of the Fermi surface at the reciprocal lattice points (111) and (200) of crystalline Al~\cite{Horsfield:1993p3925}.

We compare the analytic model to DTF-MD calculations. In our DFT-MD results, the coupling between electrons with different wavevectors is not considered. As such, it must be compared to the RPA dielectric model without the inclusion of LFC describing electron-hole coupling.
The DFT-MD absorption was calculated using the VASP code along with a KG calculation of the optical properties. The ion configurations for the optical properties calculations were obtained from snapshots of equilibrated molecular dynamics (MD) simulations within the framework of  finite temperature density functional theory~\cite{Mermin:1970p2362}.  The Al atoms are represented with frozen-core, Projector Augmented Wave (PAW) potentials~\cite{Blochl:1994p17953,Kresse:1999p1758}, with the 3{\it s} and 3{\it p} electrons in the valence band. The exchange and correlation energies are in the Perdew, Burke and Ernzerhof form~\cite{Perdew:1996p3865} of the generalized gradient approximation. The DFT-MD calculations were performed with the Baldereschi mean value point for sampling the Brilliouin zone~\cite{Baldereschi:1973p5212}. For the calculation of the optical properties, the Brillouin zone was sampled with the 10 irreducible $k$-points of a Monkhorst-Pack 6x6x6 grid~\cite{Monkhorst:1976p5188}. The plane wave cutoff energy for both the MD runs and the KG calculations was 330 eV. The results shown were obtained from simulations of 32 Al atoms in a triply periodic cubic box corresponding to a density of 2.7 g/cm$^3$. Test calculations with more atoms, higher order $k$-point sets, and higher plane wave cutoff energies did not significantly effect the results for photon energies above 15 eV. We have conducted DFT-MD absorption calculations up to a photon energy of 45 eV. Absorptions at higher energies are not included due to spurious features connected to the influence of the pseudopotential on the DOS as the L-edge is approached. Gaussian broadening is 0.1 eV.

\begin{figure}
\includegraphics[width=\linewidth]{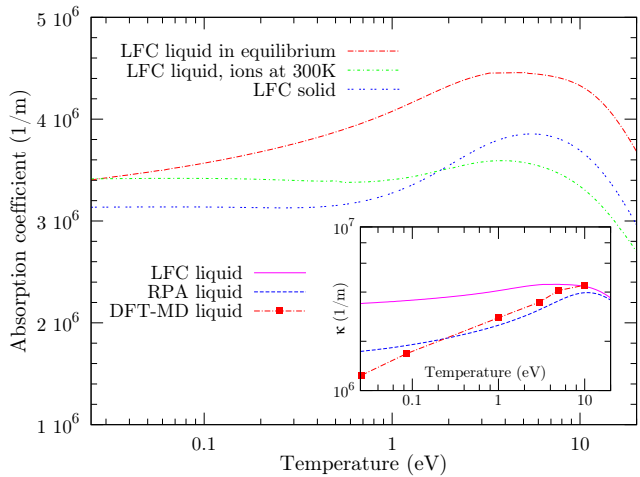}
\caption{\label{Fig:temperature} (color online) Absorption coefficient versus temperature for solid density Al, Z=3, for 30 eV photons, illustrating the absorption peak contribution of the electrons alone (green and blue) and of the combined electron-ion system (red). Inner graph shows comparison between results from Eq.~(\ref{Eq:ron-tzoar1}) in the RPA and with electron-hole LFC and DFT-MD simulations.}
\end{figure}

The absorption coefficients predicted by DFT-MD are shown in Fig.~\ref{Fig:frequency} alongside those obtained within the framework given by Ron and Tzoar both in the RPA and with LFC. Excellent agreement between the DFT-MD and RPA approaches is found in the free electron region (up to 35 eV). The discrepancy at higher energies is due to the DOS, which is no longer free-electron-like, an aspect accounted for in DFT-MD but not in the model. The LFC result differs significantly from the RPA at low temperatures but converges to it for temperatures above 10 eV, where the electronic LFC contributions become small.

An important aspect of the models is their prediction of the opacity as a function of temperature, which we show in Fig.~\ref{Fig:temperature}, for 30 eV photons. Both our analytical calculations and the DFT-MD predict a rise in the absorption coefficient up to temperatures around 10 eV. This absorption rise is due in part to plasmon peak broadening of the electron liquid and in part to the temperature broadening of the peaked, solid-like structure factor at low temperatures, as the ions heat up. Both these effects enhance the absorption since they allow more $\mathbf{k}$-vectors to participate to the process. We note that in FEL experiments the ions might not have sufficient time to thermalize and only the electron contribution is expected. For temperatures higher than the Fermi temperature, Landau damping is progressively more important with a reduction in absorption.

\begin{figure}
\includegraphics[width=\linewidth]{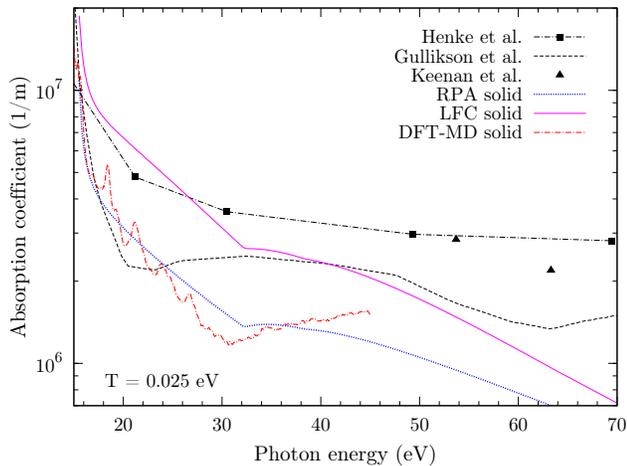}
\caption{\label{Fig:cold_solid} (color online) Absorption coefficient for solid Al. Our results with and without electron-hole coupling compared to experimental data from Henke {\it et al.}~\cite{Henke:1993p181}, Gullikson {\it et al.}~\cite{Gullikson:1994p1359} and Keenan {\it et al.}~\cite{Keenan:2002p447}.}
\end{figure}

Whilst some limited information exists about the opacity of hot dense Al~\cite{Wolfrum:2001p565}, the error bars on the experimental data are sufficiently large to nullify the usefulness of comparison with the calculations presented here.  Interestingly, even for solid Al at room temperature significant uncertainty in the absorption coefficient still remains. There are currently two sets of oft-cited experimental data for the cold opacity, given by Henke {\it et al.}~\cite{Henke:1993p181} and Gullikson {\it et al.}~\cite{Gullikson:1994p1359}, which differ by as much as a factor two in our range of interest, as can be seen from Fig.~\ref{Fig:cold_solid}. Such differences have large implications in XUV laser and high-harmonic generation research, where Al filters are often employed. Although more recent measurements using XUV lasers conducted at photon energies of 53.7 and 63.3 eV~\cite{Keenan:2002p447} show agreement with the Henke data, the Gullikson values are still widely used~\cite{CXRO}. For the sake of completeness, we also present here calculations for the cold, solid case.

The reduction of the Ron-Tzoar formalism to describe absorption in a crystalline solid is  straight-forward. In this case, the ions form a lattice and the ion-ion structure factor $S(q)$ will only depend on those q-vectors which match the reciprocal lattice vectors $\mathbf{G}$:
$
S(\mathbf{q}) = S_{\mathrm{solid}}(\mathbf{q}) = 2 \pi^2 n_i \sum_{\mathbf{G}} \delta (\mathbf{q}-\mathbf{G}).
$
Inserting this into Eq.~(\ref{Eq:ron-tzoar1}) gives the second order term of the macroscopic dielectric function for a solid, given in terms of a discrete sum over all bands. The same expression was derived by Sturm for the study of optical properties of simple metals at $T=0$~\cite{Sturm:1982p1, Sturm:1990p6973}. At frequencies above the plasma frequency and below the L-shell threshold this is the only significant term contributing to the total absorption, the intra-band (Drude) term being negligible. 

Figure~\ref{Fig:cold_solid} shows the experimentally obtained absorption coefficient for crystalline Al at 300 K of Henke {\it et al.}, Gullikson {\it et al.} and Keenan {\it et al.} in comparison with our calculations from Eq.~(\ref{Eq:ron-tzoar1}) with $S_{\mathrm{solid}}(\mathbf{q})$, with and without LFCs and with DFT-MD. We note that the RPA is inadequate to reproduce the experimental results, an observation in agreement with Sturm, who's calculations coincide with our RPA result~\cite{Sturm:1990p6973}. Good agreement is also seen between the RPA calculation and the DFT-MD at least up to the pseudo band gap around 40 eV, a result that validates the Ron-Tzoar model in the region where free-electron theory is applicable. It is immediately clear from the plot that the addition of LFC to $\varepsilon(q, \omega)$ presents a significant improvement to the calculation, placing it in rough agreement with experimental data. This is consistent with the conclusions of ref. \cite{Takada:2002p216402}. The large discrepancy between the two data sets however does not allow for a more detailed evaluation. That such discrepancies exist even in the simple cold solid case suggests the need for further accurate experimental investigations.

In conclusion, we have shown that the consideration of appropriate structure factors and potentials can be used within the framework given by Ron and Tzoar to model the free-free XUV absorption in the cold solid and that the extrapolation to the warm plasma state is consistent with DFT-MD calculations over a large range of temperatures and photon energies. Furthermore, a rise in the absorption as a function of temperature is predicted by our calculations, dictated by the behavior of both the ion and electron structure factors. Thus experiments designed to observe the temperature-dependent opacity of solid-density Al on different time scales afford the potential to substantially increase our understanding of warm dense matter.

\begin{acknowledgments}
S.M.V. gratefully acknowledges financial support from EPSRC and STFC. 
B.N. is supported by the Marie-Curie RTN `FLASH'.
M.P.D. is supported by Sandia, a multiprogram laboratory operated by Sandia Corporation, a Lockheed Martin Company, for the United States Department of Energy's National Nuclear Security Administration under Contract
No.~DE-AC04-94AL85000.
J.S.W. gratefully acknowledges support from DGAR \'Ecole Polytechnique during sabbatical.
We thankfully acknowledge discussions with Carlos Iglesias of LLNL. 
\end{acknowledgments}

\end{document}